\documentclass[11pt]{article}
\usepackage{amsmath, amssymb}

\usepackage{graphicx}

\title{Physics-Informed Neural Networks for Solving Forward and Inverse PDEs with Limited and Noisy Data: Application to Solar Corona Modeling}
\author{
  Hubert Baty \\
  Observatoire Astronomique, Université de Strasbourg, France \\
  \texttt{hubert.baty@unistra.fr}
}
\date{March 2025}

\begin{document}

\maketitle

\begin{abstract}
I will demonstrate the effectiveness of Physics-Informed Neural Networks (PINNs) in solving partial differential equations (PDEs) when training data are scarce or noisy. The training data can be located either at the boundaries or within the domain. Additionally, PINNs can be used as an inverse method to determine unknown coefficients in the equations. This study will highlight the application of PINNs in modeling magnetohydrodynamic processes relevant to strongly magnetized plasmas, such as those found in the solar corona.
\end{abstract}

\section{Introduction}

This paper builds upon the previous work of Baty \& Vigon (2024), which demonstrated the use of Physics-Informed Neural Networks (PINNs) for solving partial differential equations (PDEs) in the context of modeling magnetic fields in the solar corona. The primary objective of that study was to explore both the advantages and limitations of PINNs in comparison to traditional numerical methods, which are commonly used to solve problems with boundary conditions defined over the integration domain. In that work, the basic variant of PINNs, known as the vanilla version, was introduced. For further details, readers may also refer to Baty (2024).

In this study, we focus on scenarios where the data available at the boundaries are either highly limited and/or noisy. Furthermore, in some cases, these data may be located away from the boundary of the integration domain, for instance, at points within the domain itself. Additionally, we address inverse problems where one or more coefficients in the differential equations are unknown. These situations pose significant challenges for classical numerical methods, which may become difficult—or even impossible—to implement. As a concrete example, we examine the problem of stationary magnetic reconnection, previously studied by Baty \& Vigon (2024), where a set of incompressible magnetohydrodynamic (MHD) equations is applied in two dimensions.

This paper is structured as follows. Section 2 provides a brief overview of the fundamentals of the vanilla PINNs method. In Section 3, we examine the direct approach for problems involving sparse boundary data, noisy data, or data located outside the boundary. Section 4 introduces the inverse method for determining unknown coefficients in the PDEs. Finally, we present our conclusions in Section 5

\section{Vanilla PINNs Method}

Physics-Informed Neural Networks (PINNs) is a machine learning approach that integrates physical laws directly into the neural network training process. In its simplest form, known as vanilla PINNs, this method minimizes a loss function that combines the residuals of the differential equations and the errors from the available training data.

Consider a partial differential equation (PDE) that depends on two spatial variables \( x \) and \( y \) and written in the residual form:

\[
\mathcal{L}(u(x,y), x, y) = 0,
\]

where \( \mathcal{L} \) is the differential operator and \( u(x,y) \) is the solution to be approximated. The goal is to find an approximate function \( \hat{u}(x,y; \theta) \), represented by a neural network, that minimizes the combined loss between the data errors and the PDE residuals.
The total loss function in the vanilla PINNs method consists of two main terms:

\[
\mathcal{L}_{\text{total}} = \lambda_{\text{data}} \mathcal{L}_{\text{data}} + \lambda_{\text{PDE}} \mathcal{L}_{\text{PDE}},
\]

where \( \mathcal{L}_{\text{data}} \) is the error between the observed data values and the model predictions, and \( \mathcal{L}_{\text{PDE}} \) is the PDE residual, evaluated at a set of collocation points in the \( (x, y) \) domain. The parameters \( \lambda_{\text{data}} \) and \( \lambda_{\text{PDE}} \) are weighting factors that adjust the relative importance of the data and PDE residuals in the total loss function. 
In this work, these two factors are simply set to unity.
The two partial loss functions are evaluated using mean squared errors (see Baty 2024). Additionally, the different derivative terms (in PDE residual) are evaluated using the automatic differentiation
(AD) technique, also used for the gradient descent technique (see below).

\subsection{Gradient Descent Method for Training PINNs}

The gradient descent (GD) method is a core optimization technique used to train PINNs. It enables the efficient minimization of the loss function, which includes both the data errors and the residuals of the PDEs. The goal of gradient descent is to iteratively update the $\theta$ parameters (weights and biases) of the neural network to minimize the total loss, a measure of how well the network approximates the solution to the PDE while respecting the physical constraints.

In the context of PINNs, the total loss function typically consists of two terms: the data loss and the PDE residual loss. Gradient descent updates the network parameters based on the gradients (partial derivatives) of the loss function with respect to each parameter.
Mathematically, the update rule for a parameter \( \theta_i \) is given by:

\[
\theta_i^{(k+1)} = \theta_i^{(k)} - \text{l\_r} \frac{\partial \mathcal{L}_{\text{total}}}{\partial \theta_i^{(k)}},
\]

where:
\begin{itemize}
    \item \( \theta_i^{(k)} \) is the parameter at iteration \( k \),
    \item \( \text{l\_r} \) is the learning rate, controlling the step size of the update,
    \item \( \frac{\partial \mathcal{L}_{\text{total}}}{\partial \theta_i^{(k)}} \) is the gradient of the total loss function with respect to the parameter \( \theta_i \).
\end{itemize}

The gradients are typically computed using backpropagation, a technique that efficiently calculates the derivatives of the loss function with respect to the network parameters by applying the chain rule of differentiation. Backpropagation, combined with gradient descent, allows the network to learn from both the physical laws encoded in the PDE residuals and the available data.
The learning rate \( \text{l\_r} \) plays a crucial role in the convergence of the optimization process. If the learning rate is too large, the network may overshoot the optimal solution, leading to poor performance. Conversely, if the learning rate is too small, convergence may be slow, requiring more iterations to reach the optimal solution.

\subsection{Training data points and collocation points}

Let us consider an integration domain enclosed by a bounding boundary, denoted $\partial D$. In this study, we examine cases where the training data (represented by red dots in Figure 1) are located either on the boundary or within the integration domain. Additionally, the amount of available data may be limited and not necessarily uniformly distributed.

 \begin{figure}[!t]
\centering
 \includegraphics[scale=0.16]{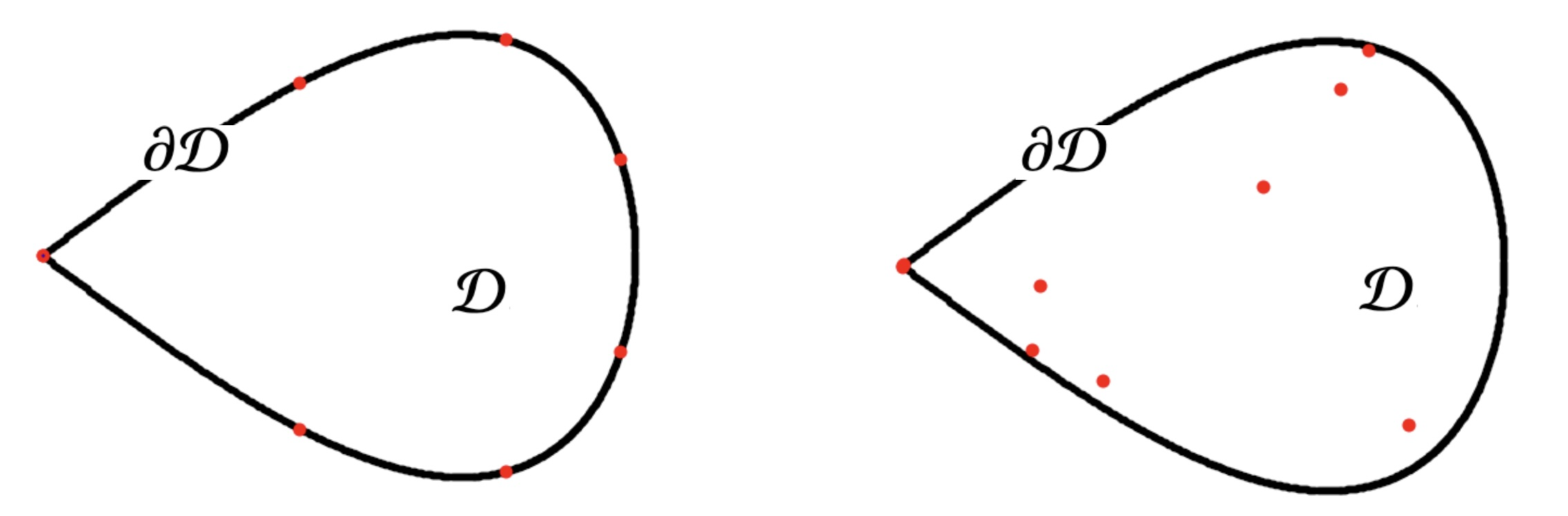}
  \caption{Schematic representation of an arbitrary 2D integration domain, and distribution of the training data points (red dots) situated on the boundary (left panel) and
  within the domain (right panel).   }
\label{fig1}
\end{figure}   

Regarding the collocation points used to evaluate the PDE residual, they are distributed throughout the domain, as illustrated in Figure 2

\begin{figure}[!t]
\centering
 \includegraphics[scale=0.46]{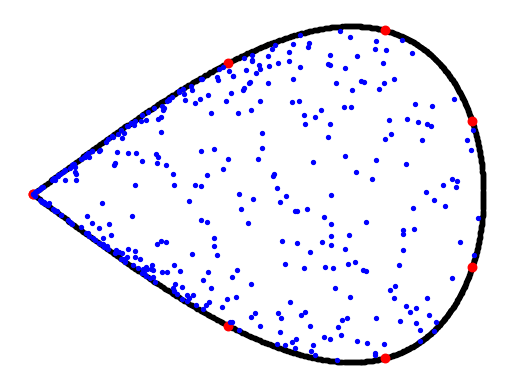}
  \caption{Schematic representation of an arbitrary 2D integration domain, and distribution of the collocation points (blue dots) within the integration domain.  }
\label{fig2}
\end{figure}   

\subsection{The neural network architecture}

The minimization is achieved using a feed-forward neural network, as illustrated in Figure 3. The network consists of an input layer with two neurons for the two spatial coordinates, followed by three hidden layers (each with five neurons), and finally by a single output neuron. In this example, only one output neuron is required when solving for a single differential equation corresponding to the $u(x, y)$ solution. The minimization process is completed when the gradient descent (GD) method converges, yielding the optimal $\theta^*$ parameter values and the associated solution.

Additionally, in the case of inverse PINNs (discussed below), extra parameters must be included in the $\theta$ parameters

\begin{figure}[!t]
\centering
 \includegraphics[scale=0.22]{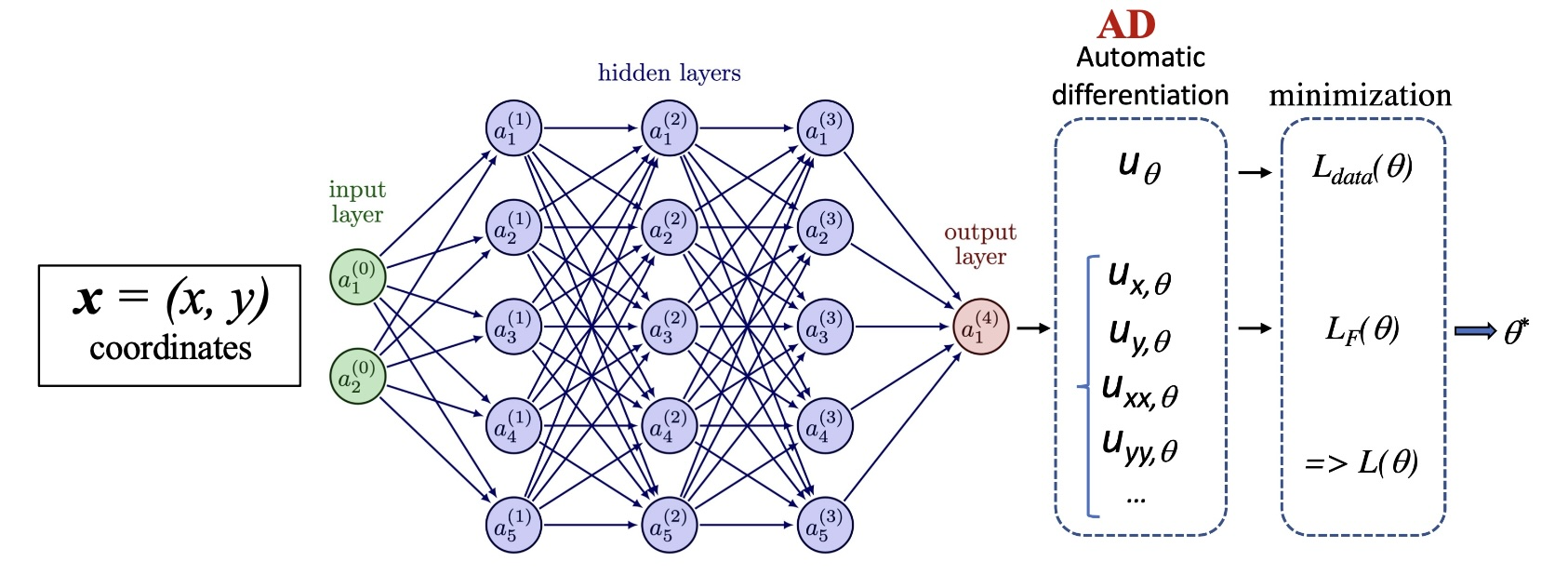}
  \caption{Example of a neural network architecture used in the PINNs method, featuring three hidden layers with five neurons per layer }
\label{fig3}
\end{figure}   

\subsection{The algorithm and implementation}

In this work, we employ the well-known Adam optimizer for the gradient descent (GD) method. The algorithm is implemented using the TENSORFLOW library, a popular Python software for machine learning. The gradient descent process is carried out with Keras, utilizing the GradientTape API. Python code will be made available on the GitHub repository: https://github.com/hubertbaty/PINNS-MHD.

\section{Direct PINNs method for MHD reconnection}

Magnetic reconnection plays a fundamental role in the release of magnetic energy during solar flares and coronal mass ejections. This mechanism, which has been extensively studied over the past 50 years, can be described within the MHD framework for incompressible, inviscid plasmas (see Priest  \& Forbes 2000 and references therein). In this context, particular 2D exact steady-state solutions exist, which are of significant interest for testing our PINNs solver. For more details on the astrophysical context, refer to Baty \& Vigon (2024) and the references therein.

\subsection{The MHD equations, spatial domain, and exact solutions}

We consider the following set of steady-state incompressible MHD equations written in usual dimensionless units (i.e.
 the magnetic permeability and plasma density are taken to be unity).
 The flow velocity obeys the equation 
 \begin{equation}
- \boldsymbol V  \cdot \nabla \boldsymbol  V + (\nabla \times \boldsymbol  B)  \times \boldsymbol  {B} - \nabla  P + \nu  \nabla^2 \boldsymbol V = 0 ,
 \end{equation}
 is written in residual form, ready to be solved by our PINNs algorithm for the two components (i.e., $V_x$ and $V_y$). Note also that, unlike the previous work (Baty \& Vigon 2024), an additional viscous term, with $\nu$ representing the viscosity coefficient, is included in this equation. This term is introduced for the sake of generality, even though the solutions we obtain are inviscid (as discussed below). The thermal pressure $P$ (through its gradient) is necessary to ensure equilibrium when solving the velocity equation. Additionally, the flow velocity vector is constrained by the incompressibility condition.
 The flow velocity vector is also constrained by the incompressibility assumption
  \begin{equation}
\nabla   \cdot  \boldsymbol  V = 0.
 \end{equation}
On the other hand, using the Maxwell-Faraday law and Ohm's law, the magnetic field vector is known to follow the equation
\begin{equation}
    \nabla \times ( \boldsymbol  V   \times \boldsymbol  B )  +  \eta \nabla^2  \boldsymbol  B = 0 ,
 \end{equation}
accompanied by the solenoidal condition
  \begin{equation}
\nabla   \cdot  \boldsymbol  B = 0 .
 \end{equation}
 Again, the residual form of the main magnetic field equation allows for the determination of the two components (i.e., $B_x$ and $B_y$). Finally, it is assumed in this work that the viscosity and resistivity coefficients, $\nu$ and $\eta$, are uniform.
 
The exact analytical solutions have been obtained by Craig \& Henton (1995), and concern a square domain
$[-1,1]^2$  in non dimensional spatial coordinates $x, y$.
They are given by the velocity and magnetic field profiles of the form:
\begin{equation}
{\boldsymbol  V } =  \left ( - \alpha x, \alpha y -   \frac  {\beta }   {\alpha }  \frac  {E_d }   {\eta \mu }   Daw(\mu x)   \right) ,
 \end{equation}
and
\begin{equation}
{\boldsymbol  B } = \left ( \beta  x   , - \beta  y  +  \frac  { E_d}  {\eta \mu   }  Daw(\mu x)   \right) ,
 \end{equation}
 where $Daw (x)$ is the Dawson function given by
\begin{equation}
Daw (x) = \int_{0}  ^{x}    \exp (t^2 - x^2)   dt    .
 \end{equation}
The definition of the $ \mu$ parameter is, 
\begin{equation}
 \mu^2 =   \frac  { \alpha^2 - \beta^2  }  {2 \eta \alpha }  ,
 \end{equation}
 where the real parameter $\beta <  \alpha$ is introduced. The real parameter $\alpha$ is generally taken to be unity, and
 $\beta$ consequently varies between $0$ and unity. 
 $E_d$ is the magnitude of a uniform electric field perpendicular to the $(x, y)$ plane which
 role is to control the rate of energy conversion.
In the limit of small resistivity $\eta$, this solution reveals a strong current sheet centered around the stagnation-point flow, with a thickness in the $x$-direction proportional to $\eta^{1/2}$. Further details about this analytical solution can be found in Baty \& Nishikawa (2016).

This analytical solution is useful for two reasons in this work. First, it provides a way to impose the solution for the training data (in terms of the two velocity and two magnetic field components) during training with gradient descent. Second, it enables the evaluation of the method's precision by directly comparing the PINNs estimate with the analytical solution at any spatial location in the integration domain.

\subsection{PINNs results for a direct solver with standard BCs }

Following the procedure outlined in Baty  \& Vigon (2024), we examine a case with $\beta = 0.75$, $E_d = 0.05$, $\alpha = 1$, and $\eta = 0.01$. Our PINNs solver must handle 6 scalar equations: the two divergence-free conditions, two scalar equations for the velocity components, and two scalar equations for the magnetic field components, along with 6 corresponding partial physics-based loss functions. Since 5 unknown variables (i.e., $V_x$, $V_y$, $B_x$, $B_y$, and $P$) define the solution, the output layer must include at least 5 corresponding neurons. In practice, we use 5 neurons, with an additional sixth neuron for the magnetic flux function $\psi$, which is used to plot magnetic field lines, as $B_x = \frac{\partial \psi}{\partial y}$ and $B_y = -\frac{\partial \psi}{\partial x}$

Regarding the network architecture, we choose 9 hidden layers, each with 30 neurons, resulting in a total of 7716 trainable parameters for $\theta$. We use $N_{data} = 40$ training data points (i.e., 10 for each boundary layer) with a random distribution, along with $N_{c} = 700$ points randomly distributed inside the integration domain. The results are obtained after a training process of 60000 epochs, using a learning rate of $l_r = 2 \times 10^{-4}$.

\subsection{PINNs results for a forward solver with limited BCs }

\begin{figure}[!t]
\centering
 \includegraphics[scale=0.42]{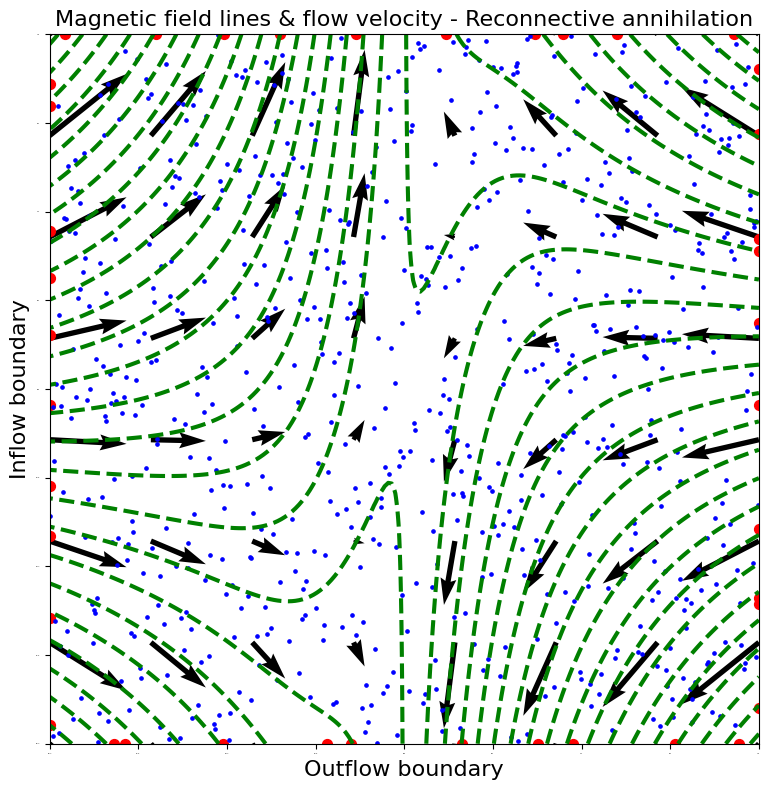}
   \caption{The solution predicted using the PINNs solver is shown, with magnetic field lines plotted as iso-contours and flow velocity represented by black arrows. The locations of the training and collocation data points are marked with red and blue dots, respectively.}
\label{fig4}
\end{figure}   

\begin{figure}[!t]
\centering
 \includegraphics[scale=0.36]{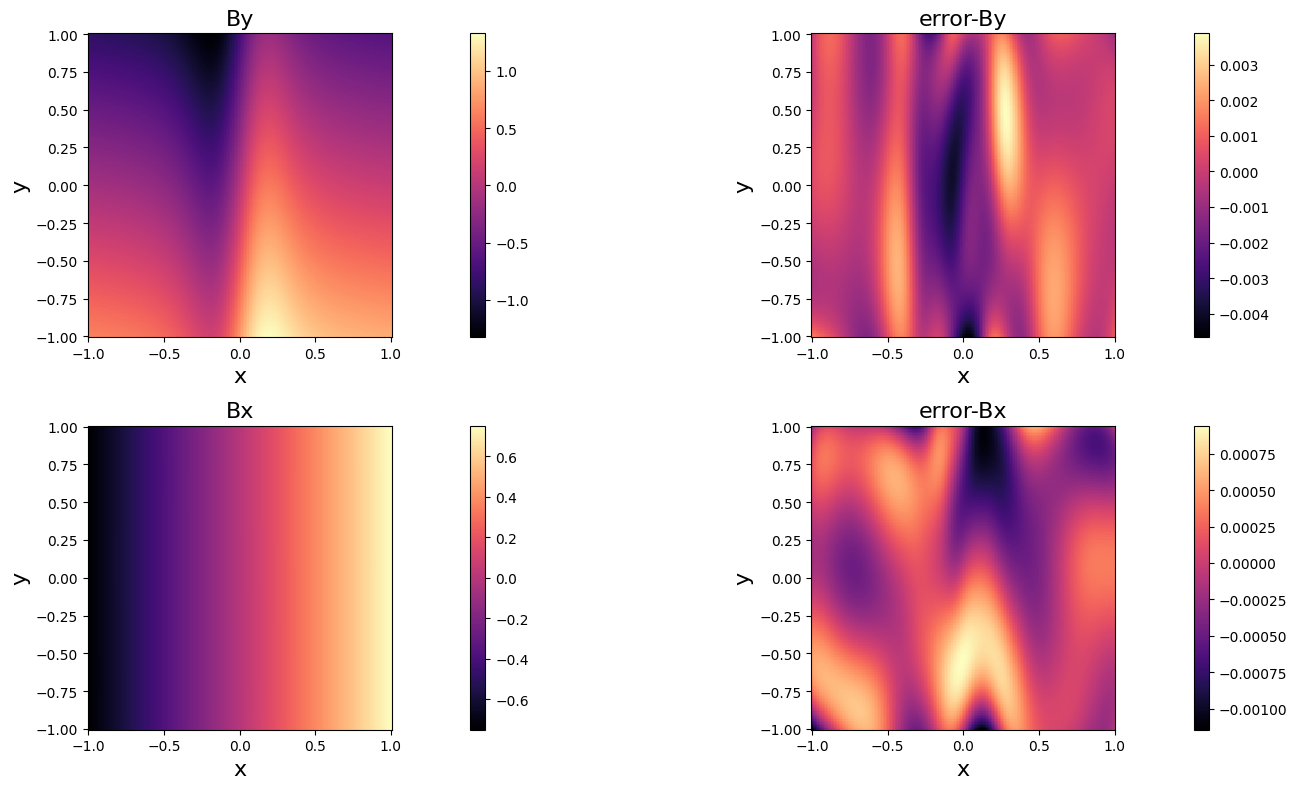}
  \caption{Colored iso-contours of the $B_y$ and $B_x$ magnetic field components predicted by PINNs solver, and associated absolute error
  distributions.}
\label{fig5}
\end{figure}   

\begin{figure}[!t]
\centering
 \includegraphics[scale=0.42]{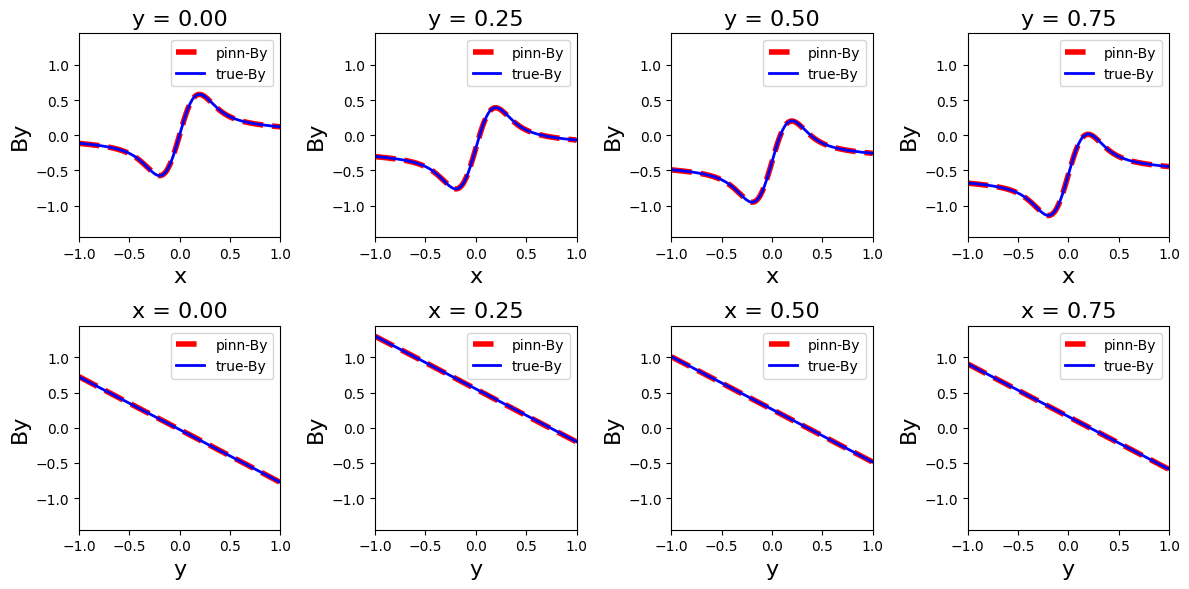}
    \caption{One-dimensional $B_y$ component (red colour) obtained for different $x$ and $y$ particular values (see legend)
  compared with the exact analytical solution (blue colour). }
\label{fig6}
\end{figure}

The solutions obtained with the PINNs solver are compared to the exact analytical solutions. The results are presented in Figs. 4-6. The maximum absolute error, on the order of $3 \times 10^{-3}$, is observed in the maps showing the spatial error distribution of the magnetic field, as seen in Fig. 5. Similar results are also obtained for the velocity components (not shown). One-dimensional cuts for different x and y values, plotted in Fig. 7, further confirm the excellent precision of the solver. These results are consistent with those obtained in Baty \& Vigon (2024), where $30$ training data points per boundary were used instead of $10$.

Note that the predicted PINNs solution and the associated error distribution are obtained using a third set of points, distinct from the collocation points. This set is taken as a uniform grid of $200 \times 200$ points.

As mentioned in the introduction, we have investigated the effect of reducing the number of training data at the boundaries. As expected, the precision decreases with fewer data points. Surprisingly, the effect becomes noticeable only when the number of data points per boundary drops below a few, and the results remain acceptable even with just 2 data points per boundary (with an error of about ten percent, as shown in Figure 8). The error is also only around 1 percent for the case with 3 data points per boundary.

\begin{figure}[!t]
\centering
 \includegraphics[scale=0.42]{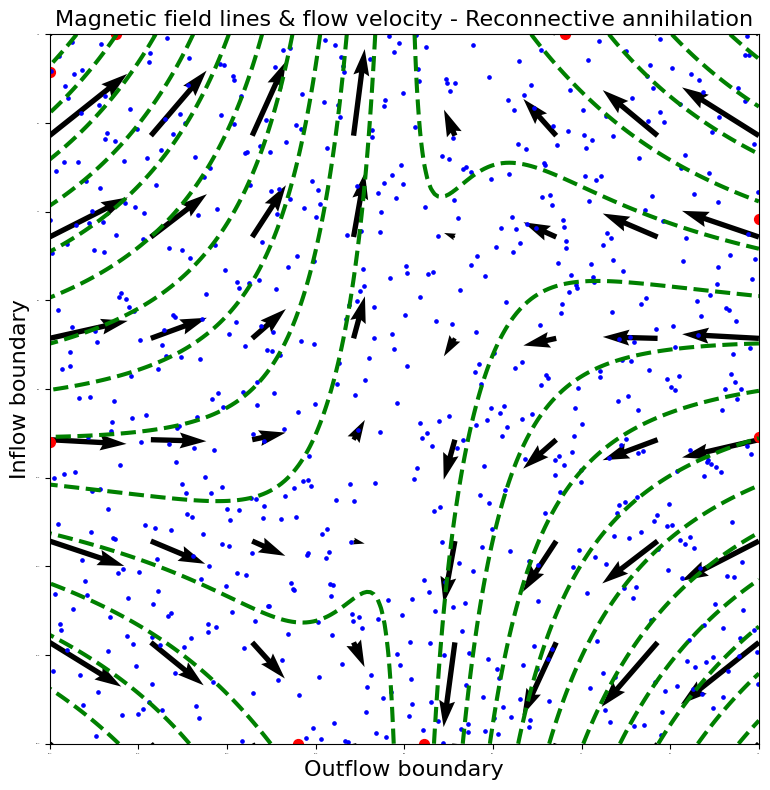}
   \caption{The solution predicted using the PINNs solver is shown, with magnetic field lines plotted as iso-contours and flow velocity represented by black arrows. The locations of the training and collocation data points are marked with red and blue dots, respectively. A set of training data with only 2 points per boundary is used.}
\label{fig7}
\end{figure}   

\begin{figure}[!t]
\centering
 \includegraphics[scale=0.36]{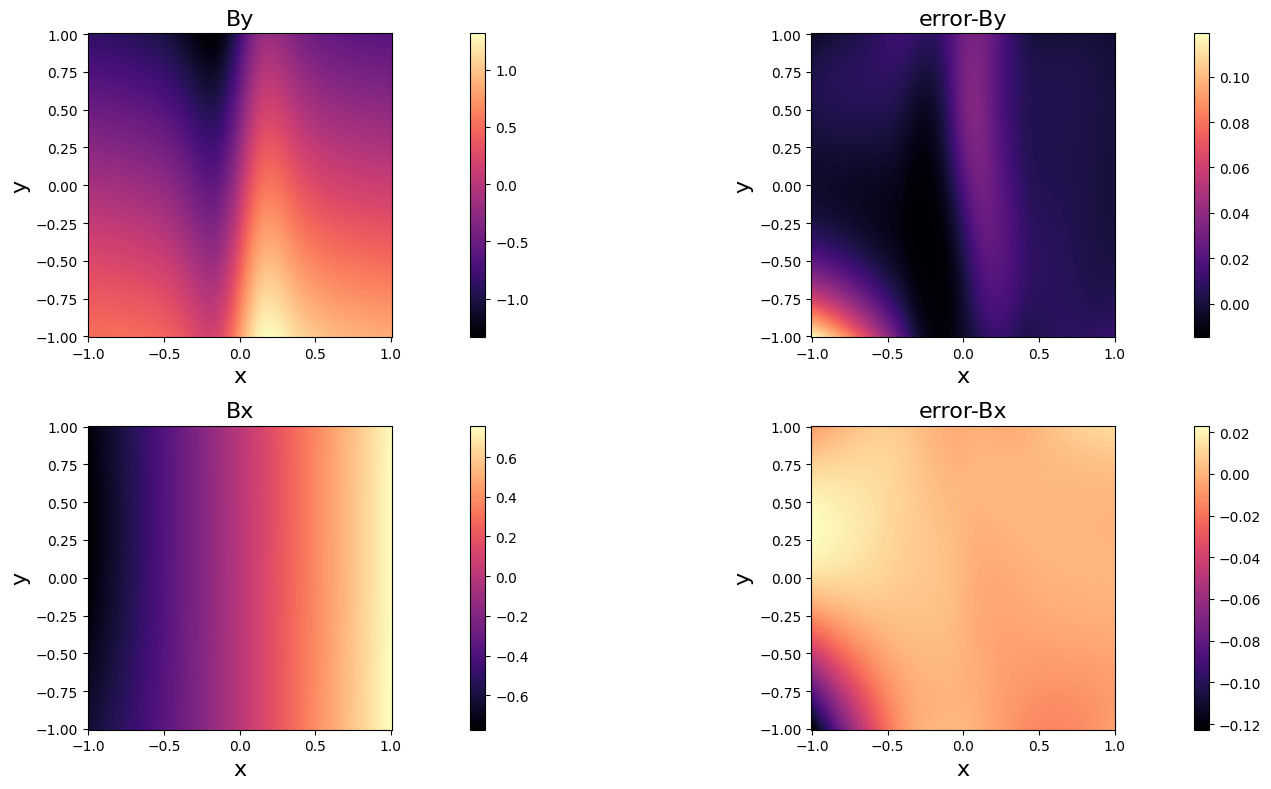}
  \caption{Colored iso-contours of the $B_y$ and $B_x$ magnetic field components predicted by PINNs solver, and associated absolute error
  distributions. This corresponds to the case of the previous figure.}
\label{fig8}
\end{figure}

\subsection{PINNs results for a forward solver with noisy BCs }

When the boundary conditions (BCs) are derived from measurements, noise can naturally affect the results. To investigate such scenarios, we introduced random noise to the exact solution used for the training data values. Our PINNs solver proved to be quite robust, as no noticeable difference was observed compared to the results obtained for the reference case using 10 data points per boundary (without noise), as long as the noise amplitude was below $10^{-3}$. However, increasing the noise amplitude leads to a degradation in precision, with the error growing more significant as the noise level increases.

\subsection{PINNs results for a forward solver with internal conditions }

As explained in the introduction, when the exact solution is known at specific locations not located at the boundary, classical integration schemes are often difficult or impossible to implement. For such problems, the PINNs solver can be used in the same way as described above, without any additional difficulty. This is demonstrated in two cases. In the first case, we used training data located within a smaller square inside the domain, and the same PINNs solver is applied to predict the solution for the entire larger square domain. In the second case, the data locations are along an arbitrary sinuous trajectory in the $y$ direction.

The results are shown in Figure 9. Regarding precision, the maximum error is on the order of a few percent and occurs in regions far from the training data, which are typically located at the edges of the integration domain.

\begin{figure}[!t]
\centering
 \includegraphics[scale=0.30]{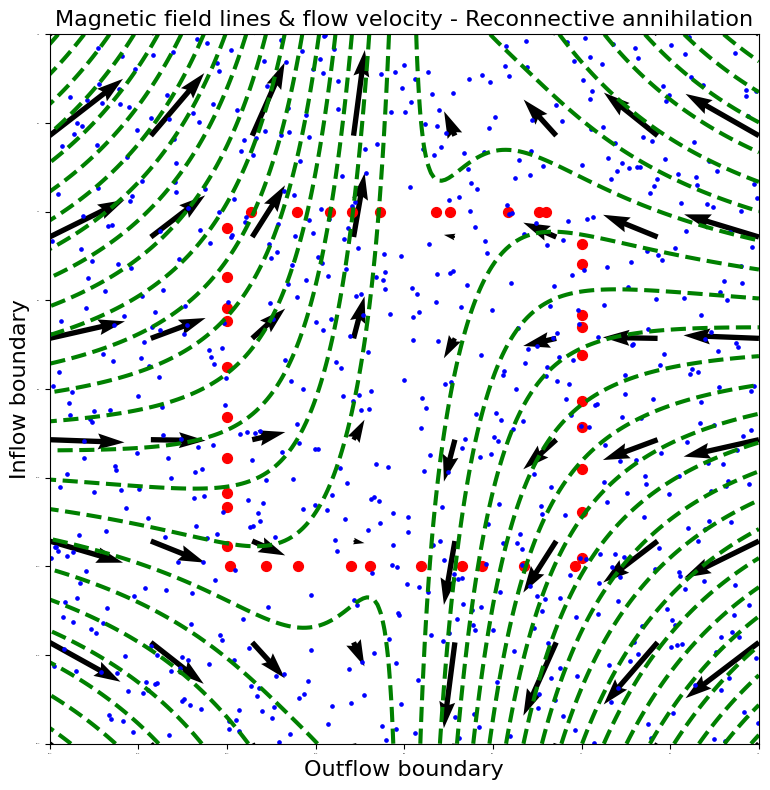}
 \includegraphics[scale=0.30]{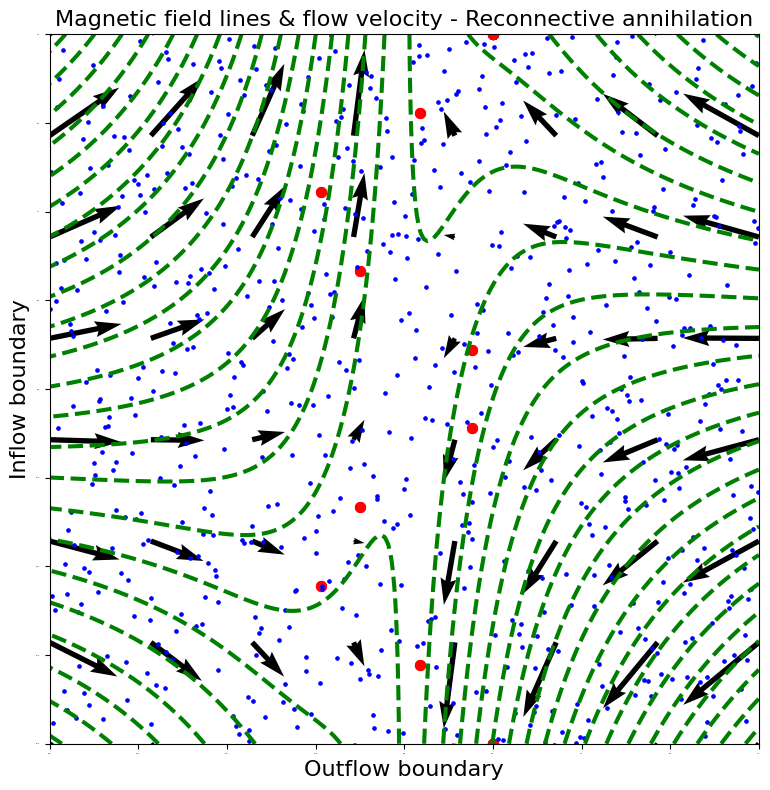}
 \caption{
 The solution predicted using the PINNs solver is shown. Magnetic field lines are plotted as iso-contours, and flow velocity is represented by black arrows. The locations of the training and collocation data points are marked with red and blue dots, respectively. (Left panel) A set of training data with 10 points per boundary is taken from a smaller square. (Right panel) A sinusoidal training data set is used.}
\label{fig9}
\end{figure}  

Finally, the facility of use PINNs with complex training data is illustrated in Figure 10, where the data depict a smiley. We can call this case smiley-driven
reconnection solution. 

\begin{figure}[!t]
\centering
 \includegraphics[scale=0.40]{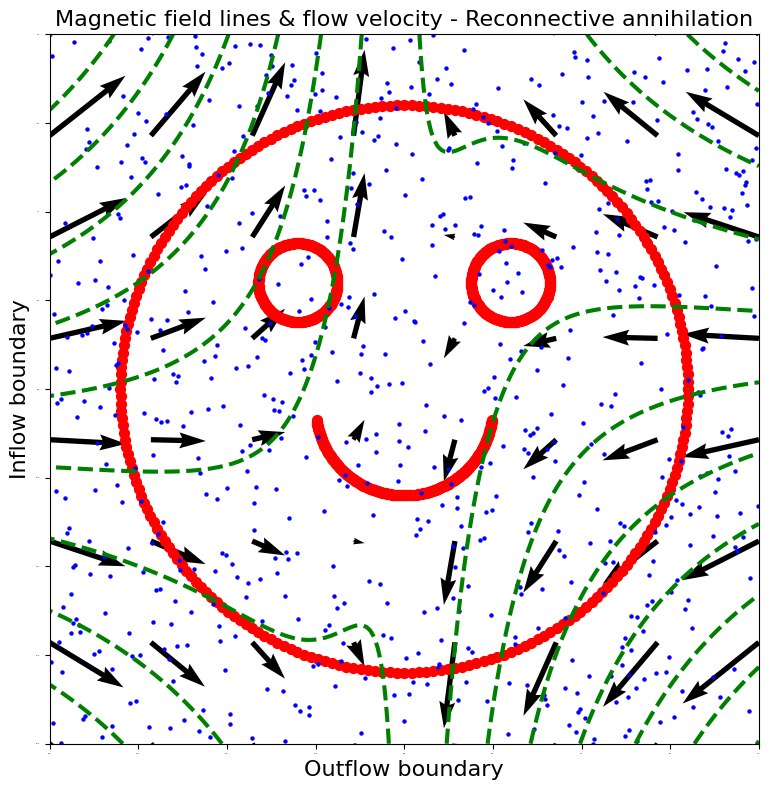}
  \caption{
  Solution predicted using the PINNs solver for a smiley-driven reconnection (see text for details).}
\label{fig10}
\end{figure}

\section{PINNs results for an inverse solver with unknown resistivity and viscosity coefficients }

For some problems, the exact values of certain coefficients in the PDEs are unknown, leading to inverse problems. In our case, this applies to the dissipation coefficients in the MHD equations, namely the resistivity and viscosity. In such situations, the PINNs method can also be used as an inverse solver. The only modification compared to the previous forward solver is the inclusion of the two coefficients in the list of $\theta$ parameters.

The results obtained for this inverse problem are similar in all respects to those reported for the standard forward problem. The convergence of the solution during training, including the values for resistivity and viscosity, is illustrated in Figure 11. Specifically, the evolution of the different loss functions as a function of the epochs (iteration number) is shown in Figure 11-a, demonstrating a similar variation to that reported for direct solvers (e.g., Baty  \& Vigon, 2024). The corresponding convergence for the two unknown coefficients, shown in Figure 11-b, indicates that the expected values of $0$ and $10^{-2}$ for viscosity and resistivity (respectively) are achieved with good precision. The maximum error in the solution is also comparable to that obtained for the direct solver.

\begin{figure}[!t]
\centering
 \includegraphics[scale=0.42]{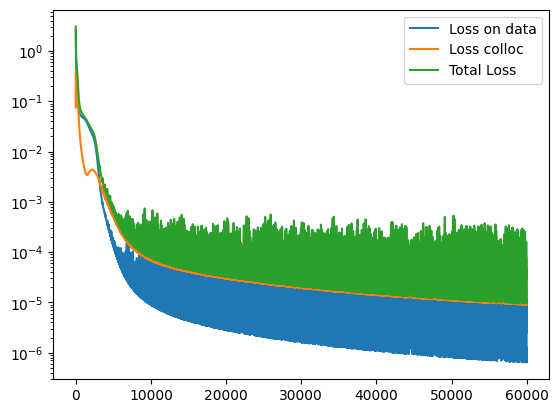}
 \includegraphics[scale=0.40]{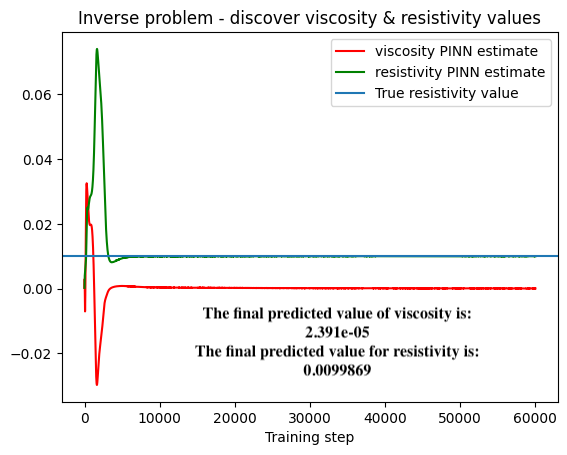}
 \caption{(a) Evolution of the loss functions over the training epochs during the training process for the inverse problem. (b) Corresponding evolution of the coefficient values (resistivity and viscosity).
 }
 \label{fig9}
\end{figure}  

\section{Conclusion}

We have used the process of steady-state MHD magnetic reconnection, relevant to the solar corona, as an interesting example to illustrate the application of the PINNs method. The corresponding set of PDEs presents a challenging numerical integration problem, particularly when the data used to define the boundary conditions are sparse and/or noisy. Another difficulty arises when these data are not available on the boundary but instead at specific locations inside the domain. In all of these cases, we have demonstrated how the same PINNs solver can be applied efficiently without introducing any additional complexity.

Moreover, another significant advantage of the PINNs method is its ability to serve as an inverse solver for determining unknown coefficients in the PDEs, without compromising the precision of the solution.

\section*{Acknowledgements}
The author thanks Vincent Vigon for fruitful discussions on PINNs technique.


\begin{thebibliography}{99}

 \bibitem[Baty \& Nishikawa 2016]{bat17} Baty H., Nishikawa H., 2016, MNRAS 459, 624
 
 \bibitem[Baty \& Vigon 2024]{bat24a} Baty H., Vigon V.,  MNRAS, 527, 2575-2584
 
  \bibitem[Baty 2024]{bat24} Baty H., 2024,  arXiv preprint arXiv:2403.00599, https://arxiv.org/pdf/2403.00599

 \bibitem[Craig \& Henton 1995]{cra95} Craig I.J.D., Henton S.M., 1995, ApJ, 450, 436 
 
  \bibitem[Priest \& Forbes 2000]{pri00}  Priest E.R., Forbes T.G., 2000, Magnetic Reconnection (Cambridge University Press, 2000)
         



\end{thebibliography}
\end{document}